\documentclass[runningheads,a4paper]{llncs}
\usepackage{amsmath}
\usepackage{amssymb}
\usepackage{subfigure}
\usepackage{graphicx}
\usepackage{multirow}
\usepackage{multicol}
\usepackage{color}

\usepackage{url}

\urldef{\mailsa}\path|{a.b, c.d, e.f, g.h|
\urldef{\mailsb}\path|i.j, k.l, m.n, o.p,|
\urldef{\mailsc}\path|emails}@urv.cat|    
\newcommand{\keywords}[1]{\par\addvspace\baselineskip
\noindent\keywordname\enspace\ignorespaces#1}

\begin{document}
\title{An Efficient Solution for Breast Tumor Segmentation and Classification in Ultrasound Images Using Deep Adversarial Learning}
\author{Vivek Kumar Singh\inst{1,\thanks{Corresponding Author: vivekkumar.singh@urv.cat}} \and Hatem A. Rashwan\inst{1} \and Mohamed Abdel-Nasser\inst{1} \and  Md. Mostafa Kamal Sarker\inst{1} \and Farhan Akram\inst{2} \and Nidhi Pandey\inst{3} \and Santiago Romani\inst{1}\and
	Domenec Puig\inst{1}}
\titlerunning{An Efficient Solution for Breast Tumor Segmentation and Classification}
\authorrunning{V.K.Singh et al.}	
\institute{DEIM, Universitat Rovira i Virgili, Spain. \and
	Imaging Informatics Division, Bioinformatics Institute, Singapore.
	\and Department of Medicine and Health Sciences, Universitat Rovira i Virgili, Spain.\\
}
\maketitle              % typeset the header of the contribution
\begin{abstract}
This paper proposes an efficient solution for tumor segmentation and classification in breast ultrasound (BUS) images. We propose to add an atrous convolution layer to the conditional generative adversarial network (cGAN) segmentation model to learn tumor features at different resolutions of BUS images. To automatically re-balance the relative impact of each of the highest level encoded features, we also propose to add a channel-wise weighting block in the network.  In addition, the SSIM and L1-norm loss with the typical adversarial loss are used as a loss function to train the model. Our model outperforms the state-of-the-art segmentation models in terms of the Dice and IoU metrics, achieving top scores of $93.76\%$ and $88.82$\%, respectively.  In the classification stage, we show that few statistics features extracted from the shape of the boundaries of the predicted masks can properly discriminate between benign and malignant tumors with an accuracy of $85\%$.

%This paper proposes an efficient solution for tumor segmentation and classification in breast ultrasound (BUS) images. The segmentation stage is based on conditional generative adversarial networks (cGAN). We propose to add an atrous convolution layer to the traditional cGAN model to learn tumor features at different resolutions of BUS images. To  automatically re-balance the relative impact of each of the highest level encoded features, we also propose to add a channel-wise weighting block in the network.  In addition, the SSIM and L1-norm loss with the typical adversarial loss are used as a loss function to train the model. Our model outperforms the state-of-the-art segmentation models in terms of the Dice and IoU metrics, achieving  top scores of $93.76\%$ and $88.82$\%, respectively.  In the classification stage, we show that few statistics features extracted from the shape of the boundaries of  the predicted masks can properly  discriminate between benign and malignant tumors with an accuracy of $85\%$. 
\end{abstract}
\section{Introduction}
Breast cancer is one of the most commonly diagnosed causes of death in women worldwide \cite{siegel2017cancer}. Screening with mammography can recognize tumors in the early stages. Despite, some breast cancers may not be captured in mammographies (e.g., in the case of dense breasts). Ultrasound has been recommended as a powerful adjunct screening tool for detecting breast cancers that may be occluded in mammographies \cite{lauby2015breast}. Computer-aided diagnosis (CAD) systems are widely used to detect, segment and classify masses in breast ultrasound (BUS) images. One of the main steps of BUS CAD systems is tumor segmentation.   \\
Over the last two decades, several BUS image segmentation methods have been proposed, which can be categorized into semi-automated and fully automated according to the degree of human intervention. In \cite{abdel2017breast}, a region growing based algorithm was used to automatically extract the regions that contain the tumors, and image super-resolution and texture analysis methods were used to discriminate benign tumors from the malignant ones. Recently, some deep learning based models have been proposed to improve the performance of breast tumor segmentation methods. In \cite{xu2019medical}, two convolutional neural network (CNN) architectures have been used to segment BUS images into the skin, mass, fibro-glandular, and fatty tissues (an accuracy of $90$\%). Hu et al \cite{hu2019automatic} combined a dilated fully convolutional network with a phase-based active contour model to segment breast tumors, achieving dice score of $88.97\%$.  \\
Although these methods and others proposed in the literature do provide useful techniques, there are still challenges due to the high degree of speckle noise present in the ultrasound images, as well as to the high variability of tumors in shape, size, appearance, texture, and location. In this paper, we propose an efficient solution for breast tumor segmentation and classification in BUS images using deep adversarial learning. \\
The main contribution of this paper is to develop an efficient deep model for segmenting the breast tumor in BUS by combining an atrous convolution network (AC) and channel attention with channel weighting (CAW) in a cGAN model in order to enhance the discriminant ability of feature representations at multi-scale. Besides, we demonstrate that the proposed segmentation model can be used for characterizing accurate shape features from the segmented mask to discriminate between benign and malignant tumors.
%\textbf{Our contributions:} 1) A new integration of channel attention with channel weighting block (CAW) has been introduced which provides a self-attention mechanism to enhance the discriminant ability of feature representations and channel weighting for BUS image segmentation, 2) An \textbf{atrous convolution block has been employed in the generator network and to consider several loss functions, such as adversarial loss (binary cross entropy),} SSIM and L1-norm, and 3) We also demonstrate that the proposed segmentation model can also be characterized by statistical descriptors on the segmented mask shape to discriminate between benign and malignant tumors.\\
The rest of this paper is structured as follows. Section 2 presents the proposed model. Section 3 includes the results. Section 4 concludes our study and provides some lines of future work.

\section{Proposed Methodology}
\begin{figure}[!h]
\centering
\includegraphics[scale=.35]{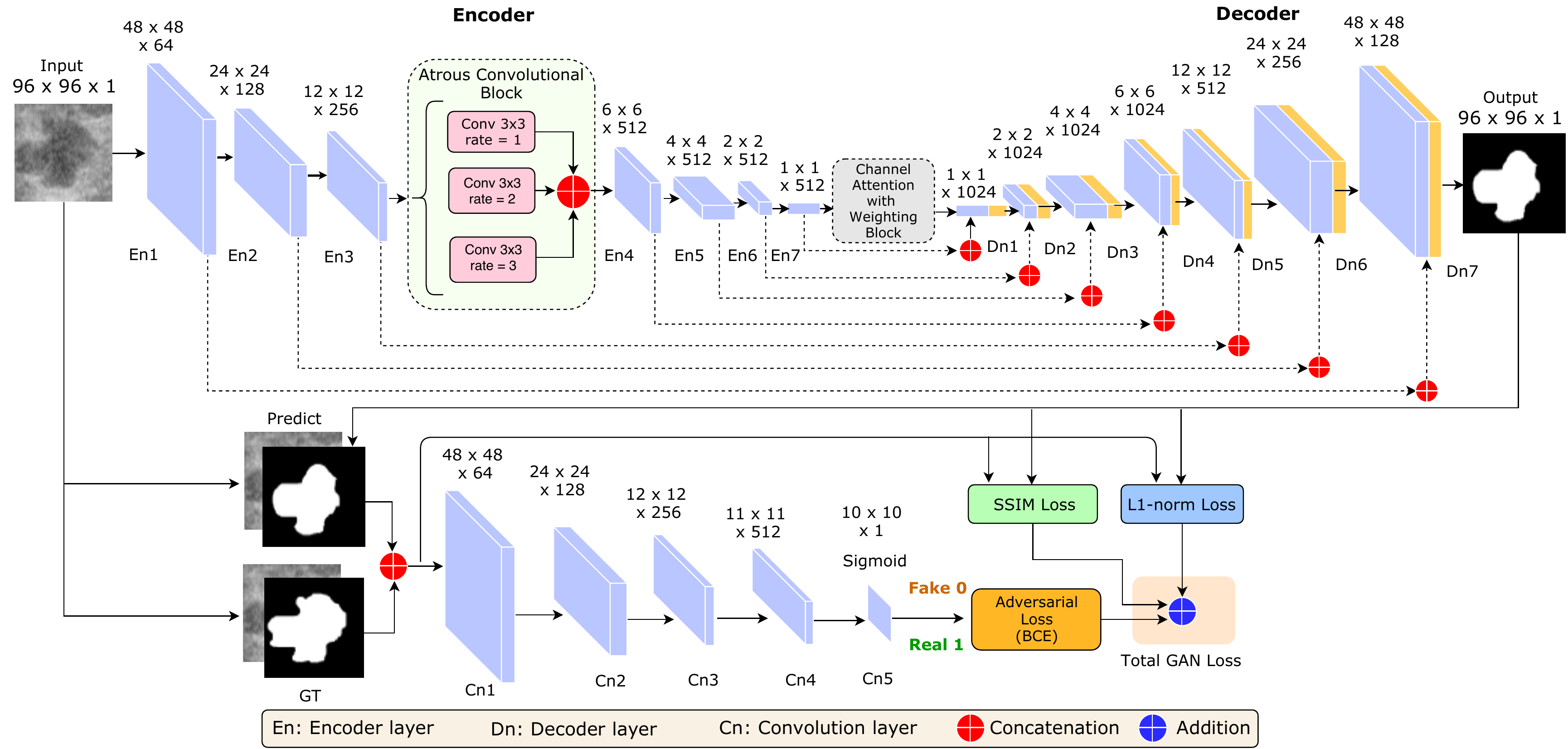}
\caption{The architecture of the proposed segmentation model for BUS images. }
\label{fig3:generator}
\vspace{-6mm}
\end{figure}

\textbf{Generative adversarial network architecture:}  The proposed BUS image segmentation technique is based on generative adversarial training, which involves two interdependent networks: a generator G and a discriminator D. (Fig. \ref{fig3:generator}). The generator generates a fake example from input noise z, while discriminator determines the probability that the fake example is from training data rather than generated by the generator.  \\
\textbf{Generator}: The generator network incorporates an encoder section, made of seven convolutional layers (En1 to En7), and a decoder section, made of seven deconvolutional layers (Dn1 to Dn7) layers. We have modified the plain encoder-decoder structure by inserting an atrous convolution block \cite{yu2015multi} between En3 and En4, in addition to a CAW block between En7 and Dn1. The CAW block is an aggregation of a channel attention module \cite{fu2018dual} with channel weighting block \cite{hu2018squeeze}. In turn, the CAW block increases the representational power of the highest level features of the generator network, which turns out in a clear improvement of the accuracy of the breast tumor segmentation in ultrasound images (for more details, see suppl. A.1 and A.2).  \\
By including the atrous convolutional block in-between encoder layers En3 and En4, the generator network is enabled to characterize features at different scales and also to expand the actual receptive field of the filters. As a consequence, the network is more aware of contextual information without increasing the number of parameters or the amount of computation. We use 1, 6 and 9 dilation rates with kernel size $3 \times 3$ and a stride of 2. \\
Each layer in the encoder section is followed by batch normalization (except for En1 and En7) and \textit{LeakyReLU} with slope 0.2, except for En7, where the regular non-linearity \textit{ReLU} activation function  is used. The decoder section is a sequence of transposed-convolutional layers followed by batch normalization, dropout with rate $0.5$ (only in Dn1, Dn2, and Dn3) and \textit{ReLU}. The filters of the convolutional and deconvolutional layers are defined by a kernel of $4\times4$ and they are shifted with a stride of 2. We add padding of 2 after En4, yielding a $4\times4\times512$ output feature map. We also add skip connection between the corresponding layers in the encoder and decoder sections, which improve the features in the output image by merging deep, coarse, semantic information and simple, fine, appearance information. After the last decoding layer (Dn7), the \textit{tanh} activation function is used as a non-linear output of the generator, which is trained to generate a binary mask of the breast tumor. \\
\textbf{Discriminator:} It is a sequence of convolutional layers applying kernels of size $4\times4$ with a stride of 2, except for Cn4 and Cn5 where the stride is 1. Batch normalization is employed after Cn2 to Cn4. \textit{LeakyReLU} with slope 0.2 is the non-linear activation function used after Cn1 to Cn4, while the sigmoid function is used after Cn5. The input of the discriminator is the concatenation of the BUS image and a binary mask marking the tumor area, where the mask can either be the ground truth or the one predicted by the generator network. The output of the discriminator is a $10\times10$ matrix having values varying from 0.0 (completely fake) to 1.0 (real). \\
\textbf{Loss Functions:} Assume $x$ is a BUS image containing a breast tumor, $y$ is the ground truth mask of that tumor within the image, $G(x, z)$ and $D(x, G(x,z))$ are the outputs of the generator and the discriminator, respectively. The loss function of the generator $G$ comprises three terms: adversarial loss (binary cross entropy loss), L1-norm to boost the learning process, and SSIM loss \cite{wang2004image} to improve the shape of the boundaries of segmented masks:
\begin{equation}
\begin{split}
 \ell_{Gen}(G, D) =  \mathbb{E}_{x,y,z}(-\log(D(x, G(x,z))))+ \\
\lambda \mathbb{E}_{x,y,z}(\ell_{L1}(y, G(x,z)))+ \alpha 
\mathbb{E}_{x,y,z}(\ell_{SSIM}(y, G(x,z)))   
\end{split}
\end{equation}
where $z$ is a random variable and $\lambda$ and $\alpha$ are empirical weighting factors. The variable $z$ is introduced as a dropout in the decoding layers $Dn1$, $Dn2$ and $Dn3$ at both training and testing phases, which helps to generalize the learning processes and avoid overfitting. If the generator network is properly optimized, the values of $D(x, G(x,z))$ should approach $1.0$, meaning that discriminator cannot distinguish generated tumor masks from ground truth masks, while L1 and SSIM losses should approach to $0.0$, indicating that every generated mask matches the corresponding ground truth both in overall pixel-to-pixel distances (L1) and in basic statistic descriptors (SSIM). For more details and analysis of loss functions, the reader is referred to suppl. A.3. \\
The loss function of the discriminator $D$ can be formulated as follows:
\begin{align}
\label{Ldis}
\ell_{Dis}(G, D) &= \mathbb{E}_{x,y,z}(-\log(D(x, y))) + \mathbb{E}_{x,y,z}(-\log(1-D(x, G(x,z))))
\end{align}
The optimizer will fit $D$ to maximize the loss values for ground truth masks (by minimizing $-\log(D(x, y))$) and minimize the loss values for generated masks (by minimizing $-\log(1-D(x, G(x, z))$). These two terms compute BCE loss using both masks, assuming that the expected class for ground truth and generated masks are $1$ and $0$, respectively. $G$ and $D$ networks are optimized concurrently: one optimization step for both networks at each iteration, where $G$ tries to generate a valid tumor segmentation and $D$ learns how to differentiate between the synthetic and real segmentation.\\
\textbf{Model training:}  In the preprocessing step, each BUS images is rescaled to 96x96 pixels, and pixel values are normalized between [0,1]. In the postprocessing step, morphological operations ($3\times3$ closing, $2\times2$ erosion) are used to suppress most of the outlier predictions (speckled pixels). The hyperparameters of the model were experimentally tuned. We also explored several optimizers, such as SGD, AdaGrad, Adadelta, RMSProp, and Adam with different learning rates (see suppl. A.3). We achieved the best results with Adam optimizer ($\beta_1$= 0.5, $\beta_2$= 0.999) and learning rate =0.0002 with a batch size of 8. The SSIM loss and L1-norm loss weighting factors $\lambda$ and $\alpha$ were set to 10 and 5, respectively. The best results were achieved by training both generator and discriminator from scratch for 40 epochs. \\
\textbf{Breast Tumor Classification: } To classify the BUS image into benign and malignant, we propose to rely on  statistic features of the segmented tumor mask to discriminate between both classes. Malignant breast tumors and benign lesions have different shape characteristics: the malignant lesion usually is irregular, speculated, or microlobulated. However, benign lesion mainly has smooth boundaries, round, oval, or macrolobulated shape \cite{yang2008measuring}.\\
In the classification method, each BUS image is fed into the trained generative network to obtain the boundary of the tumor, and then we compute 13 statistical features from that boundary: fractal dimension, lacunarity, convex hull, convexity, circularity, area, perimeter, centroid, minor and major axis length, smoothness, Hu moments (6) and central moments (order 3 and below). We implemented an \textit{Exhaustive Feature Selection} (EFS) algorithm to select the best set of features. The EFS algorithm indicates that the fractal dimension, lacunarity, convex hull, and centroid are the 4 optimal features. The selected features are fed into a Random Forest classifier, which is later trained to discriminate between benign and malignant tumors. 

\section{Experiments and Discussion}
\vspace{-2mm}
\textbf{BUS dataset.} We evaluated the performance of the proposed model using the Mendeley Data BUS dataset, which is publicly available \cite{paulo2017}. This dataset contains 150 malignant and 100 benign tumors contained in BUS images. To train our model, we randomly divided the dataset into the training set ($70\%$), a validation set ($10\%$) and testing set ($20\%$). The dataset does not have a ground truth for tumor segmentation. Thus, cooperative experts have manually segmented the tumors appearing in the BUS images.\\
\textbf{Data augmentation:} To augment the current set of available examples, we applied different operations: 1) scale the images by factors varying from 0.5 to 2 with steps of 0.25, 2) apply gamma correction on the BUS images by factors varying from 0.5 to 2.5 with steps of 0.5, and 3) flip and rotate the images. These operations yield 8K BUS images. \\
\textbf{Breast tumor segmentation results:} In Table~1, we compare the baseline cGAN model \cite{isola2017image} with three variations of our model: cGAN with atrous convolution (cGAN+AC), cGAN with channel attention and weighting (cGAN+CAW), and cGAN with AC and CAW (cGAN+AC+CAW). All of these variations are also compared  with six state-of-the-art image segmentation methods: FCN \cite{long2015fully}, UNet \cite{ronneberger2015u} SegNet\cite{badrinarayanan2017segnet}, ERFNet \cite{romera2018erfnet}, and DCGAN \cite{kim2017learning}. All methods are evaluated both quantitatively and qualitatively. For the quantitative analysis, we calculate the accuracy (ACC), Dice (DIC), Intersection over Union (IoU), sensitivity (SEN) and specificity (SPE) metrics. As shown in Table~1, the added AC and CAW blocks improves the results of the baseline cGAN model.  In addition, our model (cGAN+AC+CAW) outperforms the rest in all metrics. It achieves Dice and IoU scores of $93.76\%$ and $88.82\%$, respectively, which are the metrics that better represent the degree of coincidence between predicted and ground truth segmentation. These two results outperform the ones from the second best model in the table, the UNet model, in $5\%$ to $6\%$ absolute points over the full range, which is quite significant taking into account their proximity to the maximum value. The SegNet and ERFNet models yield the worst segmentation results on BUS images. The results of the proposed model have also been compared with other methods evaluated on different datasets. For instance, Hu et al \cite{hu2019automatic} yielded an IoU of $85.10$\% on a private BUS image dataset. In addition, Xu et al \cite{xu2019medical} achieved a Dice score of $ 89.00$\%. 

\begin{figure}
  \begin{minipage}[b]{0.45\textwidth}
    \includegraphics[trim={0 0.2cm 0 0},clip,scale=.7]{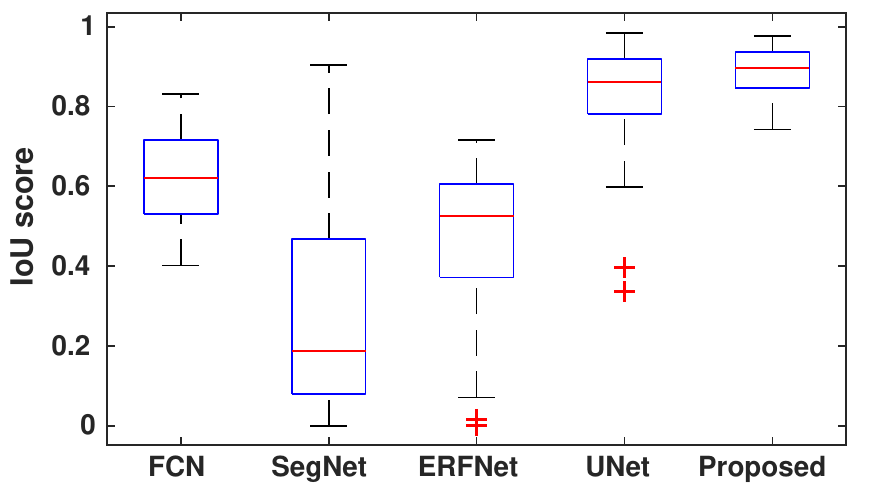}
  \end{minipage} \hspace{1cm}
  \begin{minipage}[b]{0.47\textwidth}
    \includegraphics[trim={0 0.2cm 0 0},clip, scale=.7]{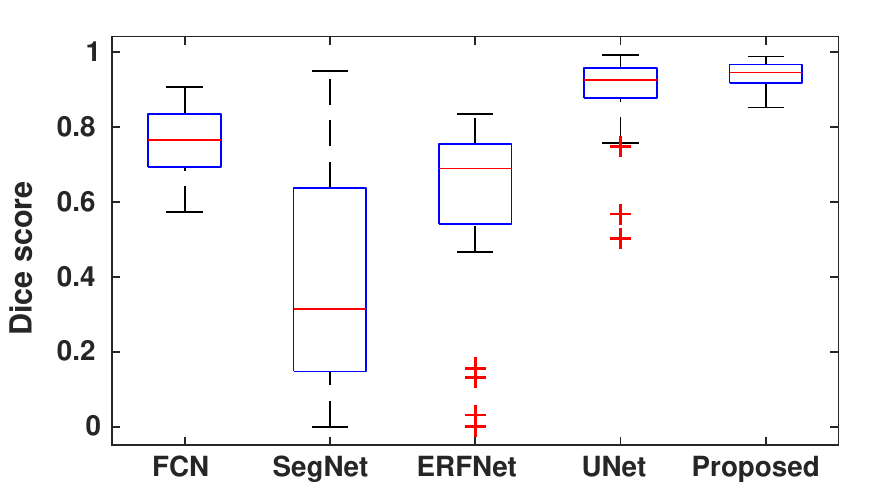}
  \end{minipage}
      
    \caption{Boxplots of IoU and Dice metrics of the proposed model and FCN \cite{long2015fully}, SegNet\cite{badrinarayanan2017segnet}, ERFNet \cite{romera2018erfnet} and UNet \cite{ronneberger2015u}. }
    \label{dice_iou}
    \vspace{-6mm}
\end{figure}

\begin{figure}[!h]
\centering
\includegraphics[scale=.72]{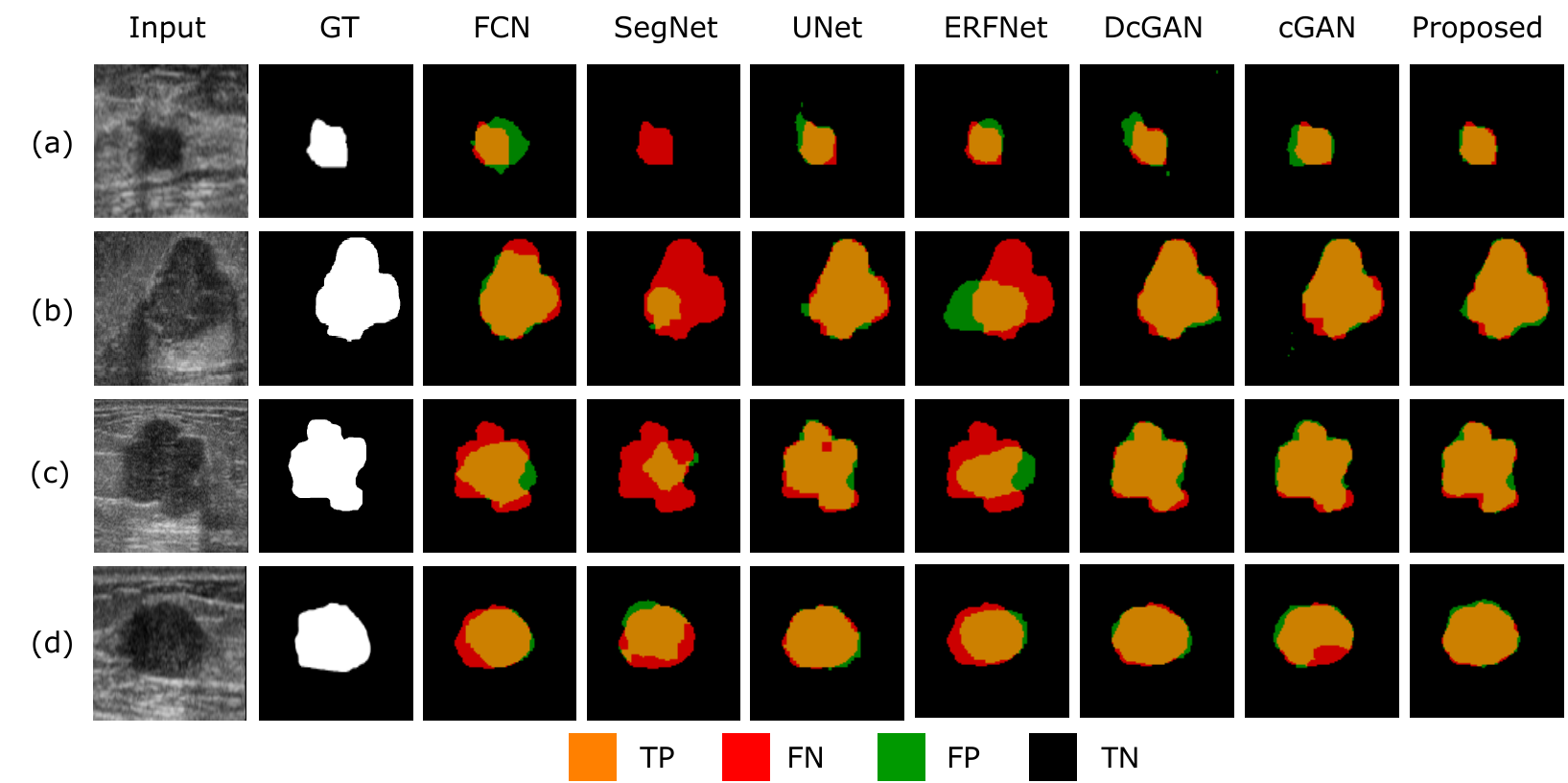}
\caption{Segmentation results on four samples of the BUS dataset. The rows (a) and (b) show benign samples while rows (c) and (d) rows show malignant samples.}
\label{fig:ss}
\vspace{-4mm}
\end{figure}

\begin{table}[!h]
	\centering
	\label{Table1}
\caption{Segmentation results of the proposed model(cGAN+AC+CAW) and compared models FCN \cite{long2015fully}, SegNet\cite{badrinarayanan2017segnet}, UNet \cite{ronneberger2015u}, ERFNet \cite{romera2018erfnet}, DCGAN \cite{kim2017learning} and cGAN\cite{isola2017image}. }
\begin{tabular}{llp{2cm}lp{8cm}}
\hline\hline
            & \textcolor{blue}{{\textbf{Methods}}}  & \multicolumn{1}{c}{\textcolor{blue}{\textit{\textbf{DIC\%}}}} & \multicolumn{1}{c}{\textcolor{blue}{\textit{\textbf{IoU\%}}}} \\ \hline\hline
\multirow{8}{*} & FCN            & $79.73\pm0.102$    & $66.29\pm0.116$         \\
                            & SegNet          & $50.35\pm0.295$    & $41.65\pm0.274$       \\
                            & UNet              & $88.28\pm0.090$    & $82.23\pm0.096$       \\
                            & ERFNet           & $67.02\pm0.206$    & $53.32\pm0.185$       \\
                            & DCGAN           & $85.55\pm0.154$    & $75.28\pm0.128$       \\
                            & cGAN            & $86.04\pm0.095$    & $77.56\pm0.080$        \\\hline
                             & cGAN + AC       & $87.14\pm0.084$    & $80.77\pm0.081$     \\
                            & cGAN + CAW       & $90.65\pm0.075$   & $83.40\pm0.069$    \\
                            & \textcolor{red}{\textbf{Ours }}      & $\textcolor{red}{\textbf{93.76}\pm0.037}$    & $\textcolor{red}{\textbf{88.82}\pm0.064}$      \\\hline
\end{tabular}
\end{table}
Figure \ref{dice_iou} shows boxplots of Dice and IoU values obtained for the 50 testing samples using FCN, SegNet, ERFNet, UNet and the proposed model. The two models based on cGAN provided small ranges of Dice and IoU values. For instance, our model is in the range $88$\% to $94$\% for Dice coefficient and $80$\% to $89$\% for IoU, while other deep segmentation methods, FCN, SegNet, ERFNet and UNet show a wider range of values. Moreover, there are many outliers in the results for the segmentation based on the other baseline segmentation methods, while using our proposal there are none.
\\Figure~\ref{fig:ss} presents a comparison between image segmentation obtained with our model and the other six models. As shown, SegNet and ERFNet yield the worst results since there are large false negative areas (in red), as well as some false positive areas (in green). FCN also shows rather significant erroneous areas, although it has fairly segmented the example (b). In turn, UNet, DCGAN, cGAN provide good segmentation but our model provide more accurate segmentation of the boundary of breast tumors, although our model shows the smallest erroneous areas, from a visual inspection.
\\ \textbf{Breast tumor classification results:} To test our classification strategy, we have checked our method with different segmentation method output with the leave-one-out cross-validation technique and calculated the precision, recall, accuracy and F1-score metrics. Furthermore, we have also obtained the same metrics from the work of Lee et.al \cite{lee2018intensity}, who proposed a stack denoising autoencoder method to segment and classify breast tumors from the same BUS dataset that we use in this study. As shown in Table ~\ref{Table2}, the proposed breast tumor classification method outperforms \cite{lee2018intensity}, with a total accuracy degree of $85\%$.
%-------------------------------------------------------
\begin{table}[!h]
	\centering
	\caption{Breast tumor classification results}
\begin{tabular}{llp{1.8cm}lp{2cm}@{}}
\hline
\hline
\textbf{Methods}          & \multicolumn{1}{c}{Precision}    & \multicolumn{1}{c}{Recall}        & \multicolumn{1}{c}{Accuracy}     & \multicolumn{1}{c}{Measure} \\\hline 
FCN      & $62.0\pm0.138$  & $70.0\pm0.154$ &  $69.0\pm0.128$           & $68.0\pm0.135$      \\\hline
SegNet      & $51.0\pm0.216$ & $62.0\pm0.238$  & $53.0\pm0.200$            & $51.0\pm0.229$     \\\hline
UNet      & $70.0\pm0.098$ & $81.0\pm0.104$ &   $77.0\pm0.091$ &  $77.0\pm0.087$      \\\hline
ERFNet   &  $58.0\pm0.162$ & $66.0\pm0.205$                   &$61.0\pm0.146$ & $59.0\pm0.172$       \\\hline
DCGAN      & $71.0\pm0.098$         &  $82.0\pm0.081$      &  $75.0\pm0.105$         & $73.0\pm0.107$     \\\hline
cGAN      & $71.0\pm0.083$         & $84.0\pm0.095$                   & $78.0\pm0.079$           & $77.0\pm0.776$      \\\hline
cGAN+AC      & $73.0\pm0.060$         & $87.0\pm0.068$                   & $80.0\pm0.053$           & $81.0\pm0.059$      \\\hline
cGAN+CAW      & $74.0\pm0.052$         & $88.0\pm0.070$                   & $82.0\pm0.056$           & $82.0\pm0.054$      \\\hline
Lee et. al \cite{lee2018intensity}      & $78.0$         & $90.0$                   & $83.0$           & $83.0$      \\\hline
\textcolor{red}{\textbf{Ours }}         & $\textcolor{red}{\textbf{81.0}\pm0.021}$        & $\textcolor{red}{\textbf{92.0}\pm0.028}$                   & $\textcolor{red}{\textbf{85.0}\pm0.196}$           & $\textcolor{red}{\textbf{84.0}\pm0.024}$          \\\hline 
\end{tabular}
\label{Table2}
\vspace{-6mm}
\end{table}

% \\ \textbf{Implementation details}: The proposed model was implemented using python with Python 3.6, CUDA 8.0, cudnn7.0, Pytorch 0.4.1 running on a 64-bit Ubuntu operating system using a 3.4 GHz Intel Core-i7 with 16 GB of RAM and Nvidia GTX 1070 GPU with 8 GB of video RAM.
\vspace{-5mm}
\section{Conclusion}
\vspace{-3mm}
In this paper, we have proposed an efficient solution for tumor segmentation  and  classification  in BUS images.  We have proposed to add an atrous convolution blocks to the generator network to learn  tumor features at different resolutions of  BUS images. We also have used a channel-wise weighting block in the generator network  to automatically re-balance the relative impact of each of the highest level encoded features. Our model outperforms the FCN, SegNet, ERFNet, UNet, DCGAN and cGAN segmentation models in terms of  Dice and IoU metrics, achieving the top scores of $93.76\%$ and $88.82\%$ respectively. In the classification stage, we used four optimal statistics features extracted from  the segmented tumor masks, obtaining an accuracy of $85\%$, which is  $2 \%$  over related method that uses the same database. As future work, we will focus on adapting the proposed model to segment lesions of different organs using multi-modal medical images.

\end{document}